\documentclass[aps,prl,twocolumn,amsmath,amssymb,superscriptaddress]{revtex4-2}  

\usepackage{graphicx}
\usepackage{amsmath}
\usepackage{amssymb}
\usepackage{epstopdf}
\usepackage{color}
\usepackage{lipsum}
\usepackage{ulem}
\usepackage[dvipsnames]{xcolor}
\usepackage{array}
\usepackage{tabularx}
\usepackage{nccmath}
\usepackage{upgreek}

\DeclareGraphicsExtensions{.eps}

\newcommand{\ket}[1] {|#1 \rangle}

\newcommand*{\rom}[1]{\expandafter\@slowromancap\romannumeral #1@}

\newcommand{\LL}[1]{\textcolor{black}{#1}}

\newcommand{\Sup}[1]{$\ket{{\Uparrow}}$}
\newcommand{\Sdown}[1]{$\ket{{\Downarrow}}$}
\newcommand{\Sempty}[1]{$\ket{{\varnothing}}$}
\newcommand{\Tup}[1]{$\ket{{\uparrow \Uparrow \Downarrow}}$}
\newcommand{\Tdown}[1]{$\ket{{\downarrow \Uparrow \Downarrow}}$}
\newcolumntype{C}[1]{>{\centering\let\newline\\\arraybackslash\hspace{0pt}}m{#1}}
\newcommand{\Vup}[1]{$\vec{S}_\Uparrow$}
\newcommand{\Vdown}[1]{$\vec{S}_\Downarrow$}
\newcommand{\Vempty}[1]{$\vec{S}_\varnothing$}
\newcommand{\Vin}[1]{$\vec{S}_{in}$}
\newcommand{\Vout}[1]{$\vec{S}_{out}$}
\newcommand{\Hpol}[1]{$\ket{H}$}
\newcommand{\Vpol}[1]{$\ket{V}$}

\begin{document}
	\title{Spin Noise Spectroscopy of a Single Spin using Single Detected Photons}
	
	\author{M. Gundín}
	\thanks{These authors contributed equally to this work.}
	\author{P. Hilaire}
	\thanks{These authors contributed equally to this work.}
	\author{C. Millet}
	\thanks{These authors contributed equally to this work.}
	\affiliation{Université Paris-Saclay, CNRS, Centre de Nanosciences et de Nanotechnologies, 91120 Palaiseau, France}
	\author{E. Mehdi}
	\affiliation{Université Paris-Saclay, CNRS, Centre de Nanosciences et de Nanotechnologies, 91120 Palaiseau, France}
	\affiliation{Université Paris Cité, Centre de Nanosciences et de Nanotechnologies, 91120 Palaiseau, France}
	\author{C. Antón}
	\affiliation{Université Paris-Saclay, CNRS, Centre de Nanosciences et de Nanotechnologies, 91120 Palaiseau, France}
	\affiliation{Depto. de Física de Materiales, Instituto Nicolás Cabrera, Instituto de Física de la Materia Condensada, Universidad Autónoma de Madrid, 28049 Madrid, Spain.}
	\author{A. Harouri}
	\author{A. Lemaître}
	\author{I. Sagnes}
	\affiliation{Université Paris-Saclay, CNRS, Centre de Nanosciences et de Nanotechnologies, 91120 Palaiseau, France}
	\author{N. Somaschi}
	\affiliation{Quandela, 7 rue Leonard de Vinci, 91300 Massy, France}
	\author{O. Krebs}
	\author{P. Senellart}
	\affiliation{Université Paris-Saclay, CNRS, Centre de Nanosciences et de Nanotechnologies, 91120 Palaiseau, France}
	\author{L. Lanco}
	\thanks{loic.lanco@u-paris.fr}
	\affiliation{Université Paris-Saclay, CNRS, Centre de Nanosciences et de Nanotechnologies, 91120 Palaiseau, France}
	\affiliation{Université Paris Cité, Centre de Nanosciences et de Nanotechnologies, 91120 Palaiseau, France}
	\affiliation{Institut Universitaire de France (IUF), 75005 Paris, France}
	
	\begin{abstract}
		Spin noise spectroscopy has become a widespread technique to extract information on spin dynamics in atomic and solid-state systems, in a potentially non-invasive way, through the optical probing of spin fluctuations. Here we experimentally demonstrate a new approach in spin noise spectroscopy, based on the detection of single photons. Due to the large spin-dependent polarization rotations provided by a deterministically-coupled quantum dot-micropillar device, giant spin noise signals induced by a single-hole spin are extracted in the form of photon-photon cross-correlations. Ultimately, such a technique can be extended to an ultrafast regime probing mechanisms down to few tens of picoseconds.\end{abstract}
	
	\maketitle
	
	Quantum systems are inherently subject to noise arising from their coupling to an environment, which represents a challenge for quantum information processing \cite{%Loss1998, 
		Ladd2010}, communication \cite{%Childress2006,
		Gisin2007}, and sensing  \cite{Degen2017}. Spin noise spectroscopy (SNS) \cite{Aleksandrov1981}, whereby magnetic fluctuations are deduced from optical noise, has emerged in this context as a powerful tool: it allows optically probing the dynamics of atomic \cite{Crooker2004} and solid-state \cite{Oestreich2005} spins.% The quantities of interest in standard SNS are spin correlators, which are usually deduced by measuring the power spectra of fluctuating polarimetric signals \cite{Sinitsyn2016}.
	
	%Standard optical SNS typically involves the measurement of the power spectrum of the light field, which as demonstrated in the Wiener-Khinchine theorem \cite{Kogan1996} retains information on the correlation of the system. 
	
	Important advances were obtained in the last decade, including broadband SNS using pulsed lasers \cite{Muller2010, Berski2013}, heterodyne SNS detection \cite{Cronenberger2016, Petrov2018, Cronenberger2019}, two-color SNS \cite{Yang2014}, access to spin correlators beyond second-order \cite{Li2016} and beyond thermal equilibrium \cite{Li2013, Glasenapp2014, Belykh2020, Guarrera2021}. %\cite{Muller2010, Li2013,Yang2014,Belykh2020}. 
	However, the low polarimetric signals imprinted by single spins \cite{Atature2007} have limited most applications to spin ensembles%, with only few reports of SNS experiments performed on single spins \cite{Dahbashi2014,Sun2022}
	. A complementary approach, quantum noise spectroscopy based on dynamical decoupling, has emerged in parallel and was fruitfully implemented with single solid-state qubits, using microwave \cite{Bylander2011,BarGill2012,Malinowski2016} or all-optical \cite{Farfurnik2023} coherent control. Dynamical decoupling can be highly desired to extend qubit coherence times \cite{Bylander2011, BarGill2012, Malinowski2016, Stockill2016, Zaporski2023, Nguyen2023} and to filter noise frequencies \cite{Bylander2011,Yuge2011}, yet implying long sequences, limited frequency bandwidths, and the availability of  coherent control techniques providing high-fidelity gates. In practice, quantum noise spectroscopy aims at measuring how the environment fluctuates, which in turn impacts the success of dynamical refocusing attempts, depending on the applied sequences. In contrast, spin noise spectroscopy aims at directly characterizing the fluctuations of the spin qubit itself, in a potentially non-invasive way \cite{Zapasskii2013}.
	
	%A promising road for the implementation of SNS with single spins relies on the cavity enhancement of polarimetric signals \cite{Poltavtsev2014,Kamenskii2020}. Pioneering experiments, reporting spin noise spectra induced by single hole spins \cite{Dahbashi2014,Wiegand2018,Sun2022}, have indeed been obtained using single quantum dots between distributed Bragg reflectors. Though they allow reaching giant polarisation rotations with spin ensembles \cite{Giri2012,Cherbunin2015}, such planar microcavities do not provide zero-dimensional optical confinement, as required for giant polarization rotations to be obtained with a single spin \cite{Arnold2015,Androvitsaneas2019,Mehdi2023}.
	
	A promising route for the implementation of SNS with single spins relies on the cavity enhancement of polarimetric signals \cite{Poltavtsev2014, Kamenskii2020}, reported by pioneering experiments on the noise spectra induced by single hole spins in planar microcavities \cite{Dahbashi2014, Wiegand2018, Sun2022}. Giant polarization rotations beyond tens of degrees have been achieved, yet only in spin ensembles \cite{Giri2012,Cherbunin2015} as higher optical confinement is required for giant rotations to be obtained with single spins \cite{Arnold2015, Androvitsaneas2019, Mehdi2023}.
	
	%nitrogen-vacancy centers in diamond \cite{Romach2019}, superconducting qubits \cite{Sung2021}, and quantum dots (QDs) \cite{Sun2022, Farfurnik2023} by spectral decomposition of noise mechanisms. 
	%\cite{Romach2015}\cite{Wilen2021}
	%Moreover, joint implementation with dynamical decoupling techniques, to filter noise frequencies, has led to enhanced sensitivity and the improvement of noise suppression strategies to extend qubit coherence times \cite{Bylander2011, BarGill2012, Stockill2016, Zaporski2023, Nguyen2023}.
	% This technique enables the spectral decomposition of the various noise mechanisms affecting these systems, allowing their identification. Moreover, noise spectroscopy along with the development of quantum sensing techniques \cite{Maze2008, Grotz2011, Grinolds2014, Lovchinsky2016, Dwyer2022} has led to the improvement of mitigation strategies to suppress the noise mechanisms and extend the qubit coherence time \cite{deLange2010, Bylander2011, BarGill2012, Farfurnik2017, Aharon2019, Ramsay2023, Zaporski2023}.
	
	%\cite{Maze2008, Grotz2011, Grinolds2014, Lovchinsky2016, Dwyer2022} 
	%\cite{deLange2010, Farfurnik2017, Aharon2019, Ramsay2023}
	
	%This method is based in general on the measurement of spin fluctuations to obtain information about the dynamics of the system.
	In this Letter, we report on a novel approach based on the measurement of the SN signal induced by a single spin, using the detection of single photons.
	Our technique takes advantage of the giant polarization rotations induced by a positively-charged quantum dot (QD) deterministically coupled to a pillar microcavity \cite{Arnold2015, Hilaire2020, Mehdi2023}.
	We implement photon-photon cross-correlations, measured along optimized polarization bases, to demonstrate strong SNS signals induced by a single hole spin.
	%We show that all the measured cross-correlations are in agreement with a theoretical model \LL{taking into account the full system's dynamics, dominated by the hyperfine interaction between the electron-in-trion and the surrounding nuclei. By providing direct access to the spin correlators, such an approach circumvents the need to measure power spectra: this allows measuring absolute SNS signals, whose maximization is directly related to the maximization of the back action induced, on the system's density matrix, by the detection of a single photon. 
	We show that all the measured cross-correlations are in agreement with a theoretical model taking into account the full system's dynamics, dominated by the hyperfine interaction between the electron-in-trion and the surrounding nuclei. By providing direct access to the spin correlators, such an approach circumvents the need to measure power spectra, enabling the measurement of absolute SN signals. The signal strength is directly related to the measurement-induced back action on the density matrix of the system by the detection of a single photon \cite{Smirnov2017}.
	Ultimately, the proposed technique paves the way towards measuring SNS in a high frequency range, limited only by the temporal jitter of single-photon detectors, to potentially reach the $50$~GHz bandwidth.

	\begin{figure*}[t]
		\includegraphics[width=\linewidth, trim={0 15cm 0 4cm}, clip]{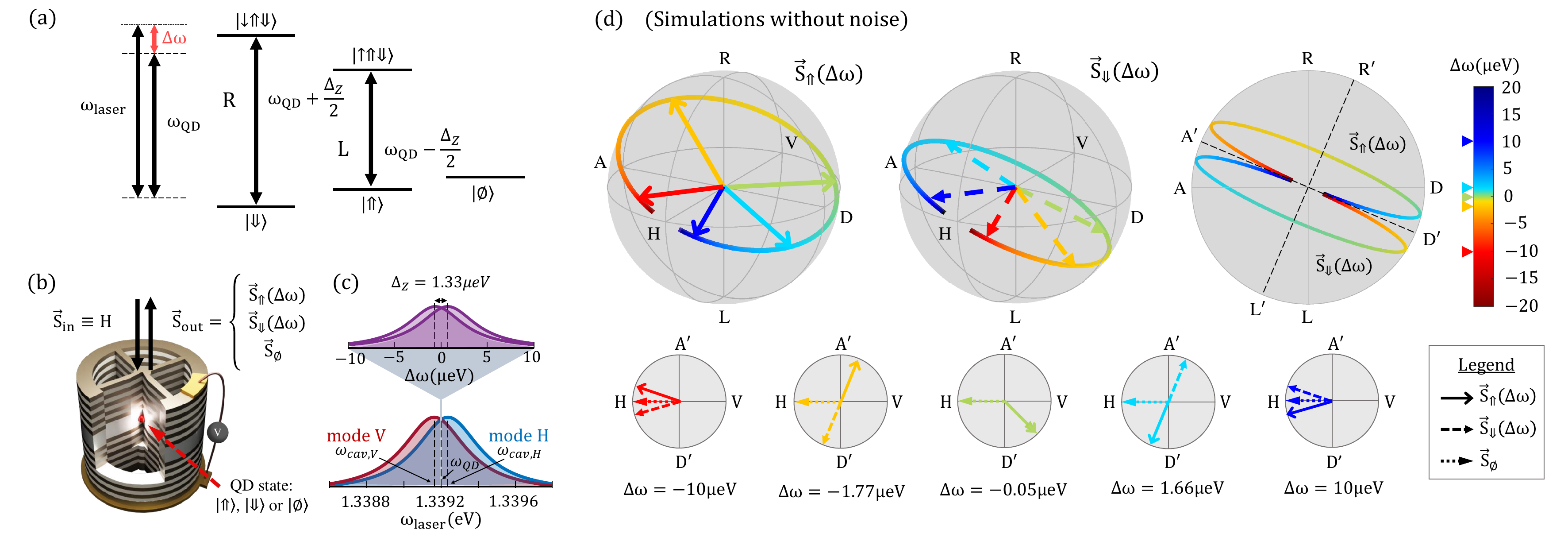}
		\caption{%\textbf{The spin-photon interface.} 
			(a) Energy level scheme of a positively charged quantum dot under a longitudinal magnetic field, inducing a Zeeman splitting $\Delta_Z$. %A H-polarized CW-laser detuned $\Delta\omega$ from the $\omega_{QD}$ transition drives the system. 
			(b) The polarization \Vout{} reflected by the micropillar depends on the QD state and the detuning of the H-polarized laser probe.
			(c) Frequency configuration of the QD transitions (top) compared with the cavity modes (bottom).
			(d) Numerical simulations, in the absence of noise, of the output polarization \Vup{} (upper left panel) and \Vdown{} (upper middle panel), as a function of $\Delta\omega$. %(see colorscale highlighting the selection of detunings for which Stokes vectors \Vup{} and \Vdown{} are displayed).
			In the upper right panel, \Vup{} and \Vdown{} are shown together, projected in a plane perpendicular to the HV axis, highlighting symmetry axes A'D' and R'L'  rotated with respect to AD and RL}.
		The lower panels show the Stokes vectors \Vup{}, \Vdown{}, and \Vempty{} for the same selected detunings\LL{, projected in the HV-A'D' plane}.
		\label{fig:polarization_rotation}
	\end{figure*}
	
	%In the following, we explain the principle of Faraday-based polarization rotation in our microcavity device, which is the core of the SNS technique that we describe afterward.
	The experiments we report are performed with the structure of Ref.~\cite{Hilaire2020}, which allows optically injecting a single hole in an annealed InAs/GaAs QD \cite{Ardelt2015}, emitting at $926~$nm.
	%In such a device, the central $\lambda$-cavity region consists of a GaAs layer that embeds the QD layer and a 20-nm-thick barrier of Al$_{0.1}$Ga$_{0.9}$As, positioned 10 nm above the QD, to drastically reduce hole tunnelling.
	%This $\lambda$-cavity is surrounded by two distributed Bragg reflectors (GaAs/Al$_{0.9}$Ga$_{0.1}$As), using 28 (14) pairs for the bottom (top) mirror, with a gradual n (p) doping for electrical contacting. %  The n-i-p structure is connected to an electrically-contacted diode through four ridges and a circular frame.
	This optical injection is implemented with a second-color, non-resonant laser at $901$~nm, with $4~\mu$W power. This allows selectively exciting the neutral exciton transition, until a photo-excited electron escapes the QD, leaving a remaining hole to populate the dot for typically $100~\mu$s \cite{Hilaire2020} (see also Supplemental Materials, hereafter referred to as \cite{SM}). %Spatial and spectral matching, between the hole-trion transition and the cavity, has been ensured through in-situ lithography \cite{Hilaire2020}.
	%, while the fine-tuning of the QD transition is enabled via Stark shift from the gate voltage. 
	
	The spin dynamics and optical properties of this device are captured by the 5-level system displayed in Fig. 1a.
	The charged QD ground states, a hole spin \Sup{} or \Sdown{}, are connected to the excited states, resp. \Tup{} or \Tdown{}, by circularly $L$ or $R$ polarized transitions \cite{Urbaszek2013}.
	%Since the hole spins form an optically inactive singlet in the excited state, we subsequently refer to the electron\LL{-in-trion} spin as $\ket{{\uparrow}}$ or $\ket{{\downarrow}}$.
	%In the absence of a magnetic field, these transitions are degenerate with energy $\omega_{QD}$ ($\hbar = 1$ units throughout the text).
	%Here, a $30$ mT longitudinal magnetic field lifts this degeneracy with a splitting \mbox{$\Delta_Z = |g_e - g_h| \mu_B B_z=1.33\,\mu$eV,} where $g_e$ and $g_h$ are the longitudinal Landé factors for the electron and hole, $\mu_B$ the Bohr magneton and $B_z$ the magnetic field along the axis.
	As in previous works on single-spin SNS \cite{Dahbashi2014,Wiegand2018,Sun2022}, a small longitudinal magnetic field, here $30$~mT, is applied to partially shield the spin from nuclear spin fluctuations. The degeneracy of the QD transition at $\omega_{QD}=1.3392$~eV ($\hbar = 1$ units throughout the text) is thus lifted with a Zeeman splitting \mbox{$\Delta_Z =1.33~\mu$eV.}
	%This magnetic field is applied to partially shield the spin from the nuclear spin fluctuations, as we discuss in detail later in this text.
	A fifth empty state \Sempty{} represents the uncharged quantum dot state. % and captures the charge tunneling dynamics.
	Both \Sup{}--\Tup{} and \Sdown{}--\Tdown{} transitions are excited by the same continuous-wave laser with energy $\omega_{\textrm{laser}}$, detuned by $\Delta\omega=\omega_{\textrm{laser}}-\omega_{\rm QD}$.

	%
	
	%As mentioned above, this is the key to this technique. We make use of numerical simulations to study this effect in the absence of noise, 
	%In order to extract information on the spin dynamics from the measurement of single photons we need a spin-photon mapping i.e. a bijective relation between the spin state and the photon polarization state.
	
	The micropillar cavity sketched in Fig. 1b has two fundamental cavity modes $\rm M= H, V$, corresponding to orthogonal linear polarizations hereafter defining the ``horizontal'' and  ``vertical'' directions.
	The incoming light field polarization is described in the Poincaré sphere by the input Stokes vector $\Vec{S}_{\rm in}$, which is chosen along the cavity eigenaxis H.
	This ensures that the reflected polarization, described by the output Stokes vector $\Vec{S}_{\rm out}$, would remain unrotated in absence of interaction with the QD optical transitions. We respectively denote \Vup{}, \Vdown{} and \Vempty{} the output Stokes vectors obtained conditionally to the system states \Sup{}, \Sdown{} and \Sempty{}  (Fig. 1b).
	%The birefringence of the micropillar cavity is small, as the mode energies $\omega_{\textrm{cav},H}$ and $\omega_{\textrm{cav},V}$ are separated $74\pm 5 \,\mu$eV compared to the mode damping rates $\kappa_H=420\pm 20 \,\mu$eV and $\kappa_V=430\pm 20 \,\mu$eV.
	%The QD, slightly red detuned $1.6\,\mu$eV from the central energy $(\omega_{\textrm{cav},H}+\omega_{\textrm{cav},V})/2$, is coupled identically to both modes.
	%Furthermore, the small Zeeman splitting $\Delta_Z$, compared with the cavity energy scale means that both QD transitions couple the same (?).
	The small birefringence of the micropillar cavity, with mode energies $\omega_{\rm cav,H}$ and $\omega_{\rm cav,V}$ separated $74\pm5~\mu$eV compared to the mode damping rates $\kappa_{\rm H}=420\pm20~\mu$eV and $\kappa_{\rm V}=430\pm 20~\mu$eV, ensures that the QD, slightly red-detuned $1.6~\mu$eV from the central energy $(\omega_{\rm cav,H}+\omega_{\rm cav,V})/2$, is coupled identically to both modes (Fig. 1c).
	These parameters, along with the cavity output coupling efficiency $\eta_{\rm top}=0.89\pm0.03$ for both modes, are extracted from polarisation-resolved experiments probing the device optical response when the system is in state \Sempty{} \cite{Hilaire2018, SM}.
	%In the case the QD is uncharged, the light does not interact with the system, thus $\Vec{S}_{\varnothing} = \Vec{S}_{\rm in} = \ket{H}$. In contrast, the interaction with a charged QD leads to a polarization rotation that depends on the detuning $\Delta\omega$ between the input field $\omega_{\rm laser}$ and the quantum dot energy $\omega_{\rm QD}$.
	
	%The dependence of the output polarizations \Vup{} and \Vdown{} with the detuning arises from an interference effect between the light emitted from the QD and the light directly reflected from the top of the cavity. 
	In the absence of environmental noise induced on the optical transitions, and in the low-power limit, \Vup{} and \Vdown{} are pure polarization states that can be analytically derived \cite{Mehdi2023,SM}. They depend on the detunings  between the laser, QD energies and cavity energies, on the cavity parameters $\kappa_{\rm H}$, $\kappa_{\rm V}$, and $\eta_{\rm top}$, but also on the QD-mode coupling strength $g$ (which plays a key role in the Purcell-enhanced emission of photons via the cavity mode), and on the rate of QD spontaneous emission in all modes other than the cavity mode, $\gamma_{\rm sp}$. In Fig. 1d we display the predicted Stokes vectors (\Vup{} in left top panel, \Vdown{} in central top panel), in absence of noise, for various
	detunings $\Delta\omega$ (see colorscale and selection of 5 specific detunings). These Stokes vectors are computed using all the previously-mentioned parameter values, together with $g=17.5\pm 0.5\,\mu$eV and $\gamma_{\rm sp}=0.9\pm 0.2\,\mu$eV, these estimations being discussed later on.
	%these two QD parameters are estimated from fits of reflectivity measurements, while keeping consistant experimental data presented later on, since they govern the shape of the device reflectivity coefficients, together with the Purcell-enhanced trion lifetime of 200~ps \cite{Ollivier2020,Suppl}. 
	%For a far-detuned excitation, the weak interaction between the input light and the QD results in a nearly unchanged output polarization (close to $H$), 
	
	As the excitation laser approaches resonance, \Vup{} and \Vdown{} experience giant rotations all around the Poincaré sphere, with a symmetric behavior highlighted in the top right panel of Fig. 1d. In this panel, the possible values of  \Vup{}($\Delta \omega$) and \Vdown{}($\Delta \omega$) are projected in the $\rm RL$-$\rm AD$ plane, with A and D antidiagonal and diagonal polarizations. It is more convenient, however, to work with the symmetry axes $\rm R'L'$ and $\rm A'D'$, which are rotated 23$^\circ$ with respect to the canonical $\rm RL$ and $\rm AD$ axes, due to the cavity birefringence \cite{SM}. In the bottom panel of Fig. 1d,  the Stokes vectors \Vup{}, \Vdown{}, and \Vempty{} are displayed as projections in the $\rm HV$-$\rm A'D'$ plane, for the five selected detunings (see legend). In such a view  \Vup{} rotates clockwise as the laser energy increases, while \Vdown{} rotates counter-clockwise.  Such a view highligts that the Stokes vectors can be drastically modified by a few $\mu$eV variation of the detuning, in the vicinity of zero detuning. It also shows an  asymmetry between \Vup{} and \Vdown{}, with respect to the HV axis,
	due to the applied $30$~mT magnetic field. These panels allow understanding why giant cross-correlations are expected, starting from the stationary regime where the spin is not initialized. For instance, at a detuning where \Vup{} and \Vdown{} are respectively close to the A' and D' polarizations, a single detected photon in polarisation A' strongly increases the conditional probability that the spin is in state \Sup{}, by Bayesian inference. This, in turn, decreases the conditional probability to detect subsequent photons in polarisation D' immediately after. 

	We now turn to the optical setup sketched in Fig. 2a. A linearly polarized CW tunable laser is sent into the pillar microcavity.
	A telescope and a cold lens inside the cryostat ensure optimal mode coupling \cite{Hilaire2018}, while a set of quarter and half waveplates align the incoming polarization with the eigenaxis of the cavity. A polarization analyzer then separates any polarization $\ket{\rm X}$ from its complementary polarization $\ket{\overline{\rm X}}$, with $\langle \rm X | \rm \overline{X}\rangle=0$, directing them to two single-photon avalanche diodes. The same setup can then be used both for the reconstruction of the output polarization states, through polarization tomography \cite{Anton2017}, and for the measurement of photon-photon cross-correlations. All undesired polarisation rotations induced by the optical setup are compensated by adjusting the angles of the various polarization waveplates.
	%, performed by scanning the CW laser across the $\omega_{QD}$ transition.
	
	%An input power of $8\,\rm pW$ is sufficiently low to ensure that the spin interacts with only one photon at a time while producing a significant SNS signal due to the cavity interaction enhancement, as well as the increase of both injection \cite{Hilaire2018} and collection \cite{Wang2019} efficiency into and from the device.
	%since photons arrive every $26 \rm ns$ on average, while the trion lifetime is $T_{rad}=170 \pm 20 \rm ps$.
	%making the probability of two photons arriving during the lifetime $\sim4.3\cdot10^{-5}$. 
	%Such a low power still produces significant output signals owing to the enhancement of the Faraday rotation provided by the micropillar cavity, as well as the increase of both injection \cite{Hilaire2018} and collection \cite{Wang2019} of photons into and from the device.
	
	\begin{figure}[t]
		\includegraphics[width=\linewidth, trim={10cm 11cm 12cm 6cm}, clip]{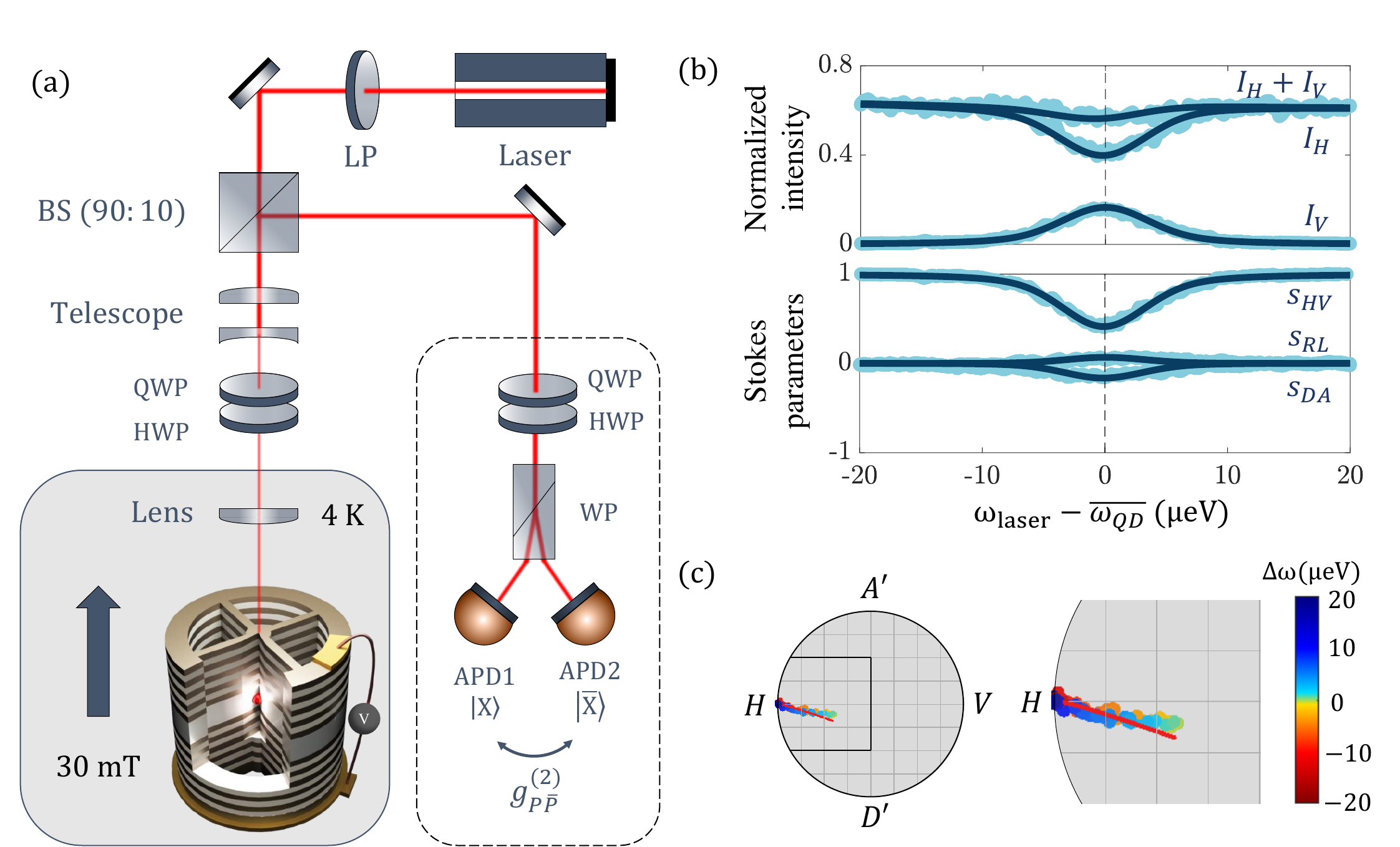}
		\caption{%\textbf{Experimental setup and output field characterization.} 
			(a) A tunable CW laser is sent into the micropillar with a polarization defined by a linear polarizer (LP). The reflected light is analyzed by a set of quarter and half waveplates (QWP and HWP) and a Wollaston Prism (WP), and each polarization measured by a set of Avalanche Photodiodes (APDs). (b) (Top) Normalized reflected intensities $I_{\rm H}$, $I_{\rm V}$, and $I_{\rm H}+I_{\rm V}$ as a function of the detuning between the laser $\omega_{\rm laser}$ and the average QD energy $\overline{\omega}_{\rm QD}$. (Bottom) Stokes parameters $s_{\rm HV}$, $s_{\rm DA}$, and $s_{ \rm RL}$ of the reflected polarization state. (c) (Left) Behavior of the reflected averaged polarization state, projected on the $\rm HV$-$\rm D'A'$ plane. (Right) Zoom on the region of interest.
		}
	\end{figure}
	
	%The output light is then analyzed in polarization with a set of waveplates and a polarizing beam splitter,  in order to map a given set of orthogonal polarizations $\ket{P}$ and $\ket{\bar{P}}$ to two single-photon detectors (see Fig. 2a).
	
	The measured output intensities along H and V polarizations, normalized by the incoming intensity, are displayed in Fig. 2b. %\LL{(Modify axis title so we see normalized intensities instead of reflectivity. Also define normalized like in the giant rotation paper, explicitely with Iin?)}
	Slow spectral fluctuations of the QD energy lead to an inhomogeneous broadening of the emission line, with average energy $\overline{\omega}_{\rm QD}$.
	%as a function of $\omega_\textrm{laser}-\overline{\omega_{QD}}$, where $\overline{\omega_{QD}}$ denotes the average QD energy taking into account inhomogeneous broadening. 
	The peak signal in the intensity $I_{\rm V}$ corresponds to the cross-polarized resonance fluorescence emitted by the QD, while the corresponding dip in $I_{\rm H}$ is the result of the destructive interference between the directly reflected laser and the co-polarized resonance fluorescence signal.
	The inhomogeneous width due to the QD spectral wandering, described by a standard deviation $\sigma_{\rm SW}=2.6\pm0.5\,\mu$eV is obtained from numerical simulations discussed later on, describing the complete QD-cavity system and taking into account the interaction of the spin with its environment \cite{SM}. This spectral wandering is large enough to induce an averaging over very different Stokes vectors, as illustrated in the lower panel of Fig. 1d. In addition, due to the lack of spin initialization, and to the limited charge occupation probability of $75\pm5\%$ (measured by a separate experiment \cite{Hilaire2020, SM}), the reflected polarization corresponds to an average of the Stokes vectors \Vup{}, \Vdown{}, and \Vempty{}.

	A complete tomography of the output state is shown in the bottom panel of Fig. 2b. The Stokes parameter $s_{\rm HV}$ is calculated as $s_{\rm HV}=\frac{I_{\rm H} - I_{\rm V}}{I_{\rm H} + I_{\rm V}}$, and analogous calculations are performed for the parameters $s_{\rm DA}$ and $s_{\rm RL}$. These provide the Stokes vector coordinates in the Poincaré sphere, as represented in Fig. 2c.
	Spectral fluctuations result in a time average of the Stokes vectors
	\Vup{} and \Vdown{}, which are no longer the pure states represented in Fig. 1 (d). This, together with the averaging of the Stokes vectors \Vup{}, \Vdown{}, and \Vempty{}, leads to a depolarization shown in Fig. 2c, where the tip of the average Stokes vector is displayed in the $HV-D'A'$ plane, and is shown to partially depolarize when $\omega_{\rm laser}$ approaches $\overline{\omega}_{\rm QD}$ \cite{SM}.

	\begin{figure*}
		\includegraphics[width=\linewidth, trim={0cm 6cm 0cm 5cm}, clip]{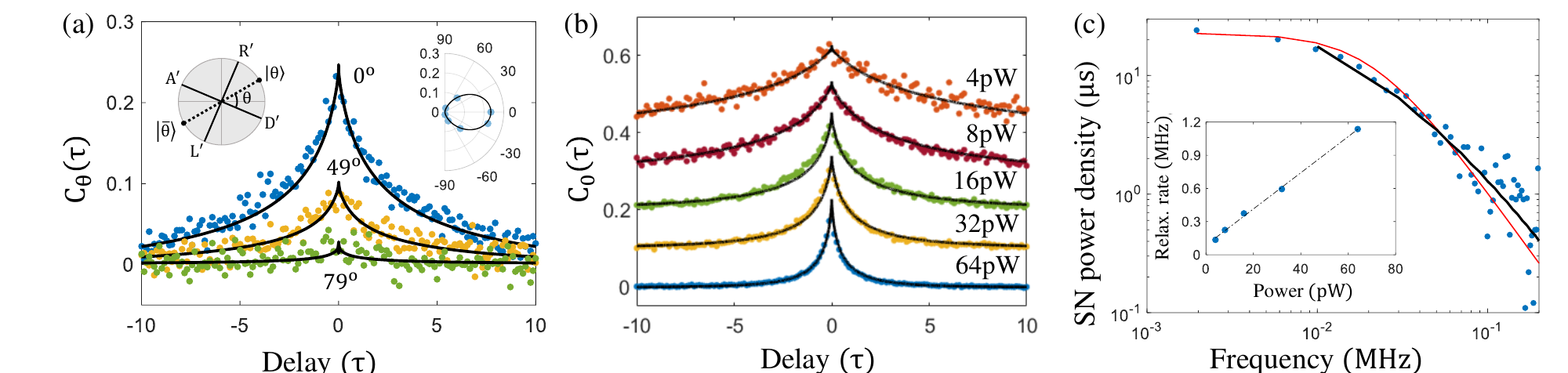}
		\caption{(a) The correlator $C_{\uptheta}(\tau)$ is measured in the basis $\uptheta\overline{\uptheta}$ for various angles (see left inset):  $\uptheta=0^\circ$ (blue dots), $\uptheta=49^\circ$ (orange dots) and $\uptheta=79^\circ$ (green dots). Right inset: maximum signal  $C_\uptheta(0)$ measured (blue dots) as a function of $\theta$. (b) The correlator $C_0(\tau)$, for a fixed angle $\uptheta=0^\circ$, is displayed for different laser powers. Each curve is displaced by $0.1$ for easier visualization. (c) SN spectrum, obtained through the FFT of the correlator $C_0(\tau)$ measured at 4 pW (blue dots : experimental data). Dashed red line : single Lorentzian fit of the SN spectrum, whose width corresponds to an effective spin relaxation rate increasing linearly with the excitation power (see inset). In all panels, solid black lines correspond to the fits obtained using a single numerical model, used to reproduce the entire set of experiments.}
	\end{figure*}
	
	%\subsection{Fig. 3}
	Intensity measurements provide an averaged picture, giving information on the system's density matrix in the stationary regime, yet conveying limited information about the system's dynamics and its intrinsic fluctuations. % \cite{Landi2023}.
	In the following, we fix the laser frequency and perform direct measurements of the output polarization fluctuations, by measuring the correlation between detection events of cross-polarized photons in the basis $\rm X\overline{X}$ separated by a varying delay $\tau$, given by the cross-correlation function:
	\begin{equation}
		g^{(2)}_{\rm X\overline{\rm X}}(\tau)=\frac{P(\overline{\rm X},\tau|\rm X,0)}{P(\overline{\rm X})},
	\end{equation}
	with $P(\overline{\rm X},\tau|\rm X,0)$ the conditional probability to detect a photon in polarization $\overline{\rm X}$ at time $\tau$, knowing that a previous photon was detected in polarization $\rm X$ at time $0$, and $P(\overline{\rm X})$ the (time-independant) probability to detect a photon in polarization  $\overline{\rm X}$. 
	
	To interpret such cross-correlations, one can first consider an ideal case where \Sempty{} is not populated, where the polarisation states \Vup{} and \Vdown{} correspond to opposite pure states in the Poincaré sphere, and where the measured polarisation bases are exactly chosen to match such states, e.g. X points along \Vup{} while $\overline{\rm X}$ points along \Vdown{}. Such a situation makes it impossible to detect a photon in polarisation X if the spin state is \Sdown{}, or to detect a photon in polarisation $\overline{\rm X}$ if the spin state is \Sup{}. Hence, a first X-polarized photon detection event indicates that the spin is in state \Sup{} immediately after detection, i.e. a perfect measurement entirely projecting the system's density matrix. All subsequent reflected photons will then be reflected in the same polarization \Vup{}, implying that no photon will be detected in the polarisation state $\overline{X}$ ($g^{(2)}_{\rm X\overline{\rm X}}(\tau) = 0$) for time delays sufficiently short compared to spin relaxation times.
	
	In practice, \Vup{} and \Vdown{} fluctuate in the Poincaré sphere due to spectral wandering, inducing random variations of the detuning $\Delta \omega$ even though $\omega_{\rm laser}$ is fixed. As seen from Fig. 1, this means that they cannot be matched to opposite detection polarizations X and $\overline{\rm X}$. In such a case, a first photon detection event in polarisation X can only create an imbalance between the spin state populations \Sup{} or \Sdown{}. This temporarily increases the probability for subsequent photons to be routed to the same detector, and decreases their probability to be routed to the other detector, measuring polarization $\overline{\rm X}$. This translates into a temporary decrease of the cross-correlation, below unity 
	($g^{(2)}_{\rm X\overline{\rm X}}(\tau)<1$), for time delays shorter than the spin relaxation time. % $\rm T_1^{\rm hole}$
	%The evolution of the system after the first photon detection event conveys valuable information about the spin dynamics.%, encoded as fluctuations in the polarization of the detected photons.
	
	In this scheme, the useful signal is %proportional to the projection of the conditional density matrix caused by the detection of the first photon. It is 
	best represented by the %cross-correlation contrast $1-g^{(2)}_{\rm X\overline{X}}(0)$.
	%We define the 
	correlator function which we generally define as $C_{\rm X}(\tau)=1-g^{(2)}_{\rm X\overline{X}}(\tau)$. This quantity %, which captures the correlations of the SNS signal,
	is strongly dependent on the chosen measurement basis $\rm X\overline{X}$. % where the Stokes vectors are projected. 
	This is shown in Fig. 3a, where the correlator $C_{\rm \uptheta}(\tau)=1-g^{(2)}_{\rm \uptheta\overline{\uptheta}}(\tau)$ is plotted for $\omega_{\rm laser}=\overline{\omega_{\rm QD}}$, for different bases 
	$\uptheta\overline{\uptheta}$ (theoretical fits will be discussed later on). All these bases are chosen in the D'A'-R'L' plane previously defined (upper right panel of Fig. 1d). As seen in the left inset of Fig. 3a, each basis $\uptheta\overline{\uptheta}$ is uniquely given by the angle $\uptheta$, measured with respect to the D'A' axis. Such a choice of the measured polarisations ensures that both $\ket{\uptheta}$ and $\ket{\overline{\uptheta}}$ are perpendicular, in the Poincaré sphere, to \Vempty{}$=\ket{H}$. As such, a detection event in polarisation $\ket{\uptheta}$ does not modify the conditional probability to be in state \Sempty{}, and only modifies the balance between the conditional probabilities to be in the spin states \Sup{} and \Sdown{}. The correlator $C_{\rm \uptheta}(\tau)$ can thus be interpreted as a spin correlator, describing the SN signal, its dynamics being only governed by spin relaxation \cite{SM}. The strength of this SN signal is described by the short-delay correlation value  $C_{\rm \uptheta}(0)$, which is maximal at $\uptheta=0$, reaching $C_0(0)=0.25$. Such a value represents a giant spin noise signal, yet lower than the maximal value of $C_0(0)=1$ which would be obtained in the ideal case of a perfect back-action (i.e. $g^{(2)}_{\rm X\overline{\rm X}}(\tau) = 0$). Notably, $C_{\rm \uptheta}(0)$
	%Such a value is directly related to the average imbalance created between spin state populations, and thus to the amount of back-action induced, on the system's density matrix, by the detection of a single photon: in comparison, a maximal value $C_0(0)=1$ would have been obtained for a perfectly projective measurement in one of the spin states  
	strongly decreases when $\uptheta$ approaches $\frac{\pi}{2}$: the dependence of $C_\uptheta(0)$, as a function of $\uptheta$, is displayed in the right inset of Fig. 3a. This dependence highlights the importance to measure along the D'A' axis ($\theta=0)$, which allows best discriminating between the polarisation states \Vup{} and \Vdown{} (see the lower panels of Fig. 1d), and thus creating the desired imbalance between the conditional occupation probabilites for states \Sup{} and \Sdown{}. Conversely, measuring along the R'L' axis ($\uptheta=\frac{\pi}{2}$) implies that $\ket{\uptheta}$ and $\ket{\overline{\uptheta}}$ are mostly perpendicular to \Vup{} and \Vdown{}, in the Poincaré sphere. A photon detected in such basis does not create a significant imbalance between the spin state populations, which translates into negligible SNS signals.
	
	After the detection event, the system's conditional density matrix
	undergoes a memory loss, induced by spin-flip processes, and thus evolves back to the steady state. This leads to a progressive decay of the correlator $C_\uptheta(\tau)$, as a function of the time delay between detection events, up to the point where such events become uncorrelated, i.e. $C_\uptheta(\tau)\xrightarrow{}0$. Fig. 3b displays the correlator measured in the optimal D'A' basis, $C_0(\tau)$, for different values of the incoming power $P_\textrm{in}$, showing that this spin relaxation becomes much faster
	at higher powers. This is typical in a positively-charged quantum dot, where the hole spin lifetime is orders of magnitude larger than the electron-in-trion spin relaxation time \cite{Dahbashi2014, Wiegand2018}. For all the incoming powers in our experiments, the trion occupation probability was large enough for spin-flip events to occur predominantly between the two trion states.
	
	Before discussing the fits in Fig. 3a and 3b, we display in Fig. 3c the spin noise spectral density, i.e. the Fourier transform of the measured correlator $C_0(\tau)$, for $P_\textrm{in}=4$~pW. The spectrum is fitted by a single Lorentzian (red solid line), following \cite{Dahbashi2014}, which allows deducing an effective spin relaxation rate from the Lorentzian Full-Width at Half-Maximum (FWHM), as $\Gamma^\textrm{(eff)}=\pi\,\times \, \rm FWHM$. This effective relaxation rate,  plotted as a function of $P_\textrm{in}$ in the inset, is approximately proportional to $P_\textrm{in}$, and thus, to the trion occupation probability. This confirms that  direct hole spin-flips between \Sup{} and   \Sdown{} play a negligible role in the effective relaxation rate. The latter is rather dominated by the electron-in-trion spin dynamics, inducing spin-flips between \Tup{} and \Tdown{}, which in turn lead to the hole spin relaxation.

	%The red solid line displays the result of a standard Lorentzian fit, which does not entirely overlap the spin noise spectrum, but still allows deducing an effective spin relaxation rate of $XX$~MHz. Furthermore, the inset of Fig. 3c highlights the fact that this effective relaxation rate, deduced from similar Lorentzian fits, is approximately proportional to $P_\textrm{in}$ (and thus, to the trion occupation probability): this confirms that the relaxation is dominated by the electron-in-trion spin dynamics. 
	
	We now turn to the numerical fits, deduced from a numerical model solving the system's master equation, and allowing to reproduce all the experiments with a single set of parameters \cite{SM}. Such a model allows computing the system's density matrix in the stationary regime (from which all fits in Fig. 2 are obtained), but also the conditional density matrix after a first photon detection in any polarisation X, and its subsequent relaxation back to the stationary regime (from which all fits in Fig. 3 are deduced) \cite{SM}. Instead of relying on the empiric description with a Lorentzian fit and an effective rate $\Gamma^\textrm{(eff)}$, this complete model also takes into account the hyperfine interaction induced by nuclear spins, in the form of a fluctuating Overhauser field \cite{Urbaszek2013}, to which the electron-in-trion coherently couples (coupling strength $\gamma_e=0.5\pm0.1\mu eV$) \cite{SM}. In such model, the electron spin precesses around an effective axis which depends on the sum of the internal Overhauser field and of the external longitudinal magnetic field (30mT). This leads, when averaging over all orientations of the Overhauser field, to a non-trivial dynamics \cite{Merkulov2002}. Following Refs \cite{Zhukov2018,Shering2019,Smirnov2020}, we also introduce in our model an additional isotropic electron spin relaxation time $\tau_e=70\pm10\rm$~ns, to take into account spin relaxation processes unrelated to hyperfine interaction \cite{SM}. This is sufficient to reproduce the dynamics of all correlators in Fig. 3a and 3b, and thus the spin noise spectra, as in Fig. 3c (black solid line), with more precision than standard empiric approaches considering single or double-Lorentzian fits of the SN spectra. In our experimental conditions, we also observed no signature of the hyperfine interaction with the hole spin, nor of any additional relaxation rate for the hole spin, which could thus remain neglected in the model \cite{,SM}.
	
	Another interest of the complete model is that, additionally to reproducing the dynamical evolution of the correlators, it allows reproducing also their absolute values, and in particular the maximal correlations $C_\theta(0)$  (Figs. 3a and 3b), which in turn govern the absolute normalization of the spin noise spectra (Fig. 3c). Fitting all these absolute values, together with fitting all the Stokes parameters and reflectivity measurements in Fig. 2d, could be obtained by adjusting 4 parameters simultaneously: the light-matter coupling strength, $g=17.5\pm 0.5\,\mu$eV, the spontaneous emission rate in all modes other than the cavity mode, $\gamma_{\rm sp}=0.9\pm 0.2\,\mu$eV, the standard deviation of the QD frequency $\sigma_{\rm SW}=2.6\pm0.5\,\mu$eV, and an additional QD pure dephasing rate $\gamma^*=0.4\pm0.1~\mu$eV \cite{SM}. We note that $g$ and $\gamma_{sp}$ are additionally constrained by the measurements of the trion relaxation rate, which is the sum of the Purcell-accelerated emission rate in the cavity mode and of the additional emission rate in all other modes, $\gamma_{sp}$ \cite{Mehdi2023,SM}. In our model,  $\sigma_{\rm SW}=2.6\pm0.5\,\mu$eV describes the effect of slow fluctuations of the QD energy, which induces slow variations of the detuning $\Delta \omega$, and thus, as seen from Fig. 1d, slow variations of the Stokes vectors \Vup{} and \Vdown{}. Such fluctuations have very different consequences from the ones induced by the pure dephasing rate $\gamma^*$, as the latter describes fast dephasing, happening within the trion radiative lifetime, which directly decreases the purity of the Stokes vectors  \Vup{} and \Vdown{} even for a fixed detuning $\Delta \omega$. While the effects of $\sigma_{\rm SW}$ and $\gamma^*$ could not be discriminated solely from reflectivity and Stokes measurements \cite{Anton2017}, they affect differently our ability to discriminate between \Vup{} and \Vdown{} with a single photon detection event, and thus can be discriminated by the additional fitting of the measured correlators. 
	
	In conclusion, we have shown that spin noise spectroscopy can also be performed with the tools of quantum optics, in the form of cross-correlations between photon detection events, taking advantage of giant polarization rotations in pillar-based structures. This approach allows directly accessing correlation functions describing the spin dynamics, and comparing them with numerical simulations computing the conditional density matrix after photon detection. The experimental data are found to fit well with a realistic model of spin noise, describing a non-trivial dynamics which can not be reproduced by phenomenological Lorentzian fits of SNS spectra. Ultimately, this technique will allow accessing much faster timescales, down to few tens of picoseconds, limited only by the temporal jitter of photon detection events. It is also the starting point for many potential studies, quantitatively addressing the phenomenon of quantum back-action induced by single photon detection events \cite{Smirnov2017, Leppenen2021}, and the related question of the entanglement between the solid-state spin and the reflected photons \cite{Hu2008, Maffei2023}.
	
	%SNS is a powerful tool for identifying the noise mechanisms that hinder the coherence time of a single spin in a quantum dot.
	%Compared to traditional SNS, we have extended this technique to the single-photon probe power regime, which allows us to directly access the correlator by performing statistics between detected photons.
	%This offers access to a broader range of timescales, from a few tens of picoseconds to hundreds of microseconds, allowing the study of ultrafast regime mechanisms.
	%Additionally, the use of spin coherent control techniques and spin initialization would greatly improve the potential of this technique. 
	
	\section*{Acknowledgements}
	We are thankful to DS Smirnov for inspiring discussions. This work was partially supported by the Paris Ile-de-France Région in the framework of DIM SIRTEQ,
	the European Union’s Horizon 2020 Research and Innovation Programme QUDOT-TECH under the Marie Sklodowska-Curie Grant Agreement No. 861097,
	the European Union’s Horizon 2020 FET OPEN project QLUSTER (Grant ID 862035), the French National Research Agency (ANR) project SPIQE (ANR- 14-CE32-0012), and a public grant overseen by the French National Research Agency (ANR) as part of the ”Investissements d’Avenir” programme (Labex NanoSaclay, reference: ANR-10-LABX-0035). This work was done within the C2N micro nanotechnologies platforms and partly supported by the RENATECH network and the General Council of Essonne.  
	
	%\section*{Author contributions}
	%P.H. and C.M. performed the experiments. M.G. analyzed the data and performed the numerical simulations. C.A. participated in the experimental developments. A.L., I.S., and N.S. fabricated the device based on a design by L.L. The project was conducted by L.L. All authors participated in scientific discussions and manuscript preparation. 

	%\bibliography{bib}
\end{document}